\documentclass[aoas]{imsart}

\RequirePackage{amsthm,amsmath,amsfonts,amssymb, multirow}
\usepackage{graphicx, xcolor}
\RequirePackage[authoryear]{natbib}

\startlocaldefs

\endlocaldefs

\begin{document}

\begin{frontmatter}
\title{SkipTrack: A Bayesian Hierarchical Model for Self-tracked Menstrual Cycle Length and Regularity in Large Mobile Health Cohorts}
\runtitle{SkipTrack}

\begin{aug}
\author[A]{\fnms{Luke}~\snm{Duttweiler}\ead[label=e1]{lduttweiler@hsph.harvard.edu}},
\author[A]{\fnms{Gowtham}~\snm{Asokan}\ead[label=e2]{gasokan@hsph.harvard.edu}},
\author[A]{\fnms{Zifan}~\snm{Wang}\ead[label=e3]{zwang@hsph.harvard.edu}},
\author[A]{\fnms{Shruthi}~\snm{Mahalingaiah}\ead[label=e4]{shruthi@hsph.harvard.edu}},
\author[A]{\fnms{Jukka-Pekka}~\snm{Onnela}\ead[label=e5]{onnela@hsph.harvard.edu}},
\author[A]{\fnms{Russ}~\snm{Hauser}\ead[label=e6]{rhauser@hsph.harvard.edu}},
\author[A]{\fnms{Michelle}~\snm{A. Williams}\ead[label=e7]{mawilliams@hsph.harvard.edu}},
\author[B]{\fnms{Kayley}~\snm{Abrams}\ead[label=e8]{kayleyabrams@apple.com}},
\author[B]{\fnms{Christine}~\snm{L. Curry}\ead[label=e9]{christine\_curry@apple.com}}
\and
\author[A]{\fnms{Brent}~\snm{A. Coull}\ead[label=e10]{bcoull@hsph.harvard.edu}}
\address[A]{Harvard T.H. Chan School of Public Health \printead[presep={,\ }]{e1,e2,e3,e4,e5,e6,e7,e10}}


\address[B]{Apple Inc. \printead[presep={,\ }]{e8,e9}}

\end{aug}

\begin{abstract}
Menstrual cycle length and regularity are important vital signs with implications for a variety of acute and chronic health conditions. Large datasets derived from cycle-tracking mobile health apps are being used to investigate the effects of various covariates on menstrual cycle length and regularity. One limitation on these analyses is that recorded cycle lengths can be incorrectly inflated if users skip tracking any cycle related bleeding days in the app. Here we present SkipTrack, a novel Bayesian hierarchical framework for examining baseline and time-varying effects on menstrual cycle length and regularity while accounting for the uncertainty of possible skips in cycle tracking. In simulations we demonstrate the superiority of the SkipTrack model by showing that competing methods which specify cycle skips a priori are more susceptible to issues of estimation bias and overconfidence than the SkipTrack model. Finally, we apply the SkipTrack framework to data from the Apple Women’s Health Study, a US-based digital cohort (consent provided at study enrollment) to examine patterns of association between age, BMI and race/ethnicity, and menstrual cycle length and regularity.
\end{abstract}


\end{frontmatter}


\section{Introduction}

Digital health plays an increasingly central role in almost every area of public health research and practice. A critical aspect of research using data from mobile devices and wearables is understanding how usage patterns of the devices impact patterns in the observed data. For example, in digital smoking intervention trials, usage patterns can serve as an important predictor of the success of the intervention [\cite{bricker2018trajectories}].  In other settings, usage patterns must be characterized to understand the representativeness of collected data, as patterns can vary by participant demographics or over time [\cite{flitcroft2020demographic}].   When interest focuses on variation in health status over time,  care must be taken to separate the biologic variation of primary interest from variation in mobile device usage over time in order to yield valid inferences.  

This work is motivated by the need to adjust for mobile health usage patterns, specifically skipping the logging of period bleeding days in menstrual tracking apps, in menstrual health research. Menstrual cycle characteristics are known to be important vital signs in menstruators, serving as an important sentinel for multiple adverse reproductive health outcomes [\cite{diaz2006menstruation, vollmar2024menstrual}]. While relatively understudied, there is an emerging body of literature establishing associations between menstrual cycle characteristics and various demographic or environmental factors [eg. \cite{cho2001effects, merklinger2017effect, giorgis2020can, hammer2020environmental, li2023menstrual}]. Recently, multiple studies have taken advantage of the massive data sets derived from mobile health menstrual cycle tracking apps to investigate factors associated with these characteristics on a large scale [eg. \cite{bull2019real,grieger2020menstrual, mahalingaiah2022design, wang2024menarche}]. These mobile health studies have some significant advantages over studies employing in-person recruitment, such as the ease of data collection and large sample sizes. 

\begin{figure}[t]
    \centering
    \includegraphics[width = .9\linewidth]{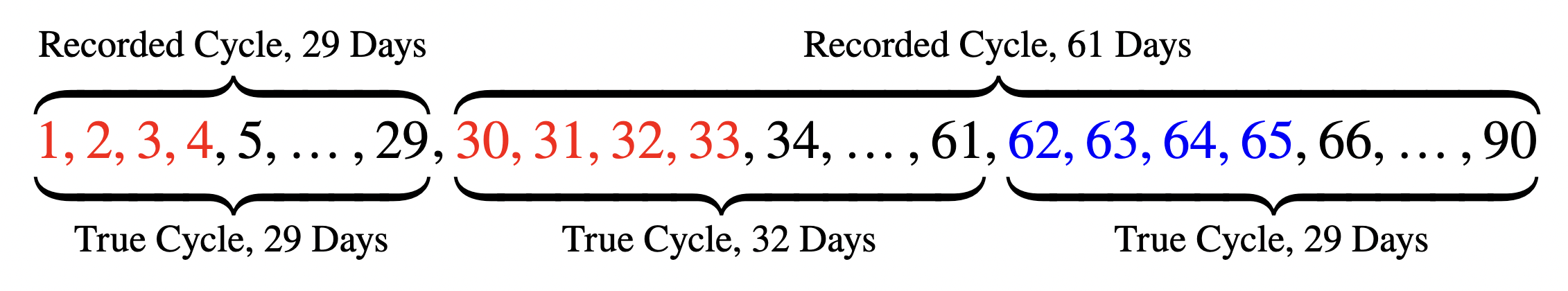}
    \caption{An example of a skip in cycle tracking. Red numbers are period days tracked by the user, blue numbers are period days not tracked by the user.}
    \label{fig: skippedTrack}
\end{figure}

Notwithstanding these advantages, the analysis of menstrual cycle data from cycle tracking apps can present challenges since the reporting of period bleeding days is self-initiated.  Mobile health apps record cycle lengths as the number of days from first day of a bleeding period, to the day prior to the first day of the next bleeding period.  If a user skips logging period days in the app, then multiple cycles incorrectly appear in the data as a single cycle, resulting in an inflated cycle length [\cite{li2022predictive}]. Figure \ref{fig: skippedTrack} provides a visualization of this phenomenon in which the red numbers represent bleeding days tracked by the user, and the blue numbers represent bleeding days which the user fails to track. 

In addition to skipped cycles in app-based tracking, another key consideration in the analysis of longitudinal menstrual cycle data is the important role that within-person cycle variability plays in reproductive health. It has been well-established that, along with fluctuations in mean cycle length, cycle irregularity is a menstrual disorder that is known to be  associated with a range of reproductive disorders, including polycystic ovary syndrome (PCOS), endometriosis, and uterine fibroids [see \cite{jukic2015lower, agarwal2019clinical, strauss2023yen}], along with cycle irregularity that naturally occurs in the perimenopause window.  Therefore, statistical analysis of menstrual cycle data should consider effects of risk factors on both the mean and within-person variance of menstrual cycle lengths. 

There is a limited amount of existing work on methods to identify skipped-tracking of menstrual cycles. To address the former problem, some large mobile health analyses have taken the basic approach of simply removing cycles outside a commonly accepted window (10-90 days), and others exclude cycles based on cycle length differences [eg. \cite{bull2019real, gibson2022covid, li2023menstrual}].
\cite{li2020characterizing} used cycle length difference (the difference in length between two consecutive cycles) and user-app interaction to remove potentially questionable cycles. Most closely related to the work we propose here, 
\cite{li2022predictive} proposed a Bayesian hierarchical model
designed to impute missed cycle tracking, and therefore predict unobserved cycle lengths contained in recorded lengths that in fact represent multiple cycles.  This model assumes a Poisson distribution for cycle lengths, thereby implicitly modeling the subject-specific mean and variance of cycle length with a single parameter,  due to the well-known equivalence of the mean and variance of a Poisson distribution.  A common denominator of all of this work, however, is that all of these approaches either exclude or estimate cycle lengths in a pre-processing step, and then fix the resulting cycle lengths as if they were observed without error in downstream association analyses.   

In this article we present SkipTrack, a novel Bayesian hierarchical model designed to estimate associations between demographics or exposures of interest and both cycle length and regularity, while accounting for skipped tracking in the cycle length data. The model is motivated by questions being addressed in the Apple Women's Health Study (AWHS), a collaboration among Harvard University, Apple Inc. and the National Institute of Environmental Health Sciences (NIEHS), designed to  gain a deeper understanding of how demographic, lifestyle and environmental factors impact menstrual cycles and gynecologic conditions including infertility, menopause, and polycystic ovary syndrome (PCOS) [\cite{mahalingaiah2022design, gibson2022covid, li2023menstrual}].  SkipTrack may be used to estimate log-linear effects on cycle length and regularity directly, to identify skipped cycles, and to rigorously quantify the uncertainty associated with each of these assessments. 

Our proposed framework differs from earlier models in a couple of important ways.  First, it does not identify skipped cycles in a pre-processing step and treat them as fixed and known in downstream regression analyses, but rather propagates uncertainty from the estimation of skipped cycles when quantifying  the uncertainty associated with regression coefficient estimates.  Second, it explicitly separates the parameterization of mean cycle length and variability, specifying a separate regression model for each.  This means that both mean cycle length and cycle length variability inform the identification of skipped cycles.  Intuitively, the degree of regularity can aid in identifying skipped cycles: it is easier to determine that a cycle recorded as 55 days results from a skipped cycle for a subject with a relatively regular pattern of 28, 30, 29, 32 days for their other cycles, whereas this determination is more uncertain for a subject with an irregular pattern of 24, 30, 46, 28 days for their other cycles.  Therefore, our approach leverages information in both the mean and variance of an individual's observed cycle pattern when identifying skipped tracking.  

In Section \ref{sec: Data}, we introduce the AWHS data motivating this work and analyzed in this paper. Section \ref{sec: Methods} presents the SkipTrack model formulation, along with details on the Markov Chain Monte Carlo (MCMC) sampler used for estimation and inference.  Section \ref{sec: Sims} presents results from a simulation study designed to evaluate the performance of the SkipTrack model in multiple data generating scenarios, and to compare this performance with that from existing methods used to handle skipped tracking.    Section \ref{sec: Application} 
presents the results from the analysis of the AWHS data, and Section \ref{sec: Discussion} discusses limitations and future research directions. 

\section{Data}\label{sec: Data}

Our work is motivated using data from the Apple Women's Health Study (AWHS). The AWHS focuses on studying factors affecting reproductive health by collecting demographic, behavioral, and menstrual cycle data through iPhone use and wearable devices, see \cite{mahalingaiah2022design} for full details of the study design. 

In short, AWHS collects survey data, manually logged menstrual cycle data, and passive wearable sensor data through the Apple Research app installed on a user's iPhone. Users are eligible if they have an iPhone, live in the United States, are at least 18 years old (19 in Alabama and Nebraska, 21 in Puerto Rico), and have provided informed consent to participate in the study. The data set used for this manuscript is comprised of data collected from November 14, 2019 (the start date for AWHS) until December 31, 2021, and includes 57,349 participants who provided data on 858,807 menstrual cycles which, after exclusions (see Section \ref{sec: Application}), resulted in an analytic dataset of 664,461 cycles contributed by 43,683 individuals.

The focus of our analysis is patterns of association between age, race/ethnicity, and BMI on both menstrual cycle length and regularity, while accounting for possible skips in cycle tracking that result from self-reported menstrual cycle start and end dates. In addition, we controlled for other self-reported covariates: smoking status, alcohol consumption, physical activity, stress levels, highest education level, self-rated socioeconomic status (SES), and parity. 

Participant age (per logged cycle) was binned into categories 20-24, 25-29, 30-34, 35-39, 40-44, 45-49, 50+, and participants were placed in the race/ethnicity categories `Asian', `Black Only', `Hispanic', `More Than One', `Other', or `White Only', based on survey responses. In our treatment of the covariate variables we followed previous work on the AWHS given in \cite{li2023menstrual}.

\section{Methodology}\label{sec: Methods}

In the following section we discuss our methodology in depth, presenting the assumed model and discussing our algorithm for estimation and inference.

\subsection{Notation and Framework}

Consider a sample of $n$ individuals, each contributing $n_i$ observed cycle lengths for a total of $N = \sum_i n_i$ observations. Let $y_{ij}$ be the $j$th observed cycle length for individual $i$, where $i = 1, \dots, n$ and $j = 1, \dots, n_i$. We occasionally reference all observations from individual $i$ with the $n_i \times 1$ vector $\mathbf{Y}_i$, and all observed cycle lengths with the $N \times 1$ vector $\mathbf{Y}$.

Let $\mathbf{X}_i$ be a $n_i \times p$ matrix of covariates for the cycle length mean of individual $i$. We use $X_{ij}$ to refer to the $p \times 1$ vector of covariates for the mean corresponding to the observation $y_{ij}$, and occasionally refer to the entire $N \times p$ matrix with $\mathbf{X}$. Then let $\mathbf{Z}$ be a $n \times q$ matrix of baseline covariates for cycle length `regularity,' generally parameterized as precision or variance. We use $Z_i$ to refer to the $q\times 1$ vector of covariates corresponding to individual $i$, or the $i$th row of $Z$. Note that both baseline and longitudinal covariates can be included in $\mathbf{X}$, while only baseline covariates are allowed in $\mathbf{Z}$.

\cite{li2023menstrual} demonstrated that baseline characteristics may have an impact on \textit{both} menstrual cycle length and regularity. Thus, we allow for the inclusion of covariates simultaneously in $\mathbf{X}$ and $\mathbf{Z}$. This can result in equivalent covariate information in $\mathbf{X}$ and $\mathbf{Z}$ if no longitudinal covariates are included.

Now, recall that each observed cycle length $y_{ij}$ may correspond to one or more biological cycle lengths. We represent this number of biological, or `true', cycles with a variable $c_{ij}$, where $c_{ij} = 1$ if $y_{ij}$ is a true cycle length, $c_{ij} = 2$ if $y_{ij}$ is the length of two true cycles, and so on. These values are unobservable in our application and framework, as well as not generally being target parameters of interest. However, including the $c_{ij}$ values as latent variables in our model allows us to eliminate sources of potential bias and inflated Type I error.

\subsection{Model}

\begin{figure}
    \centering
    \includegraphics[width = .85\linewidth]{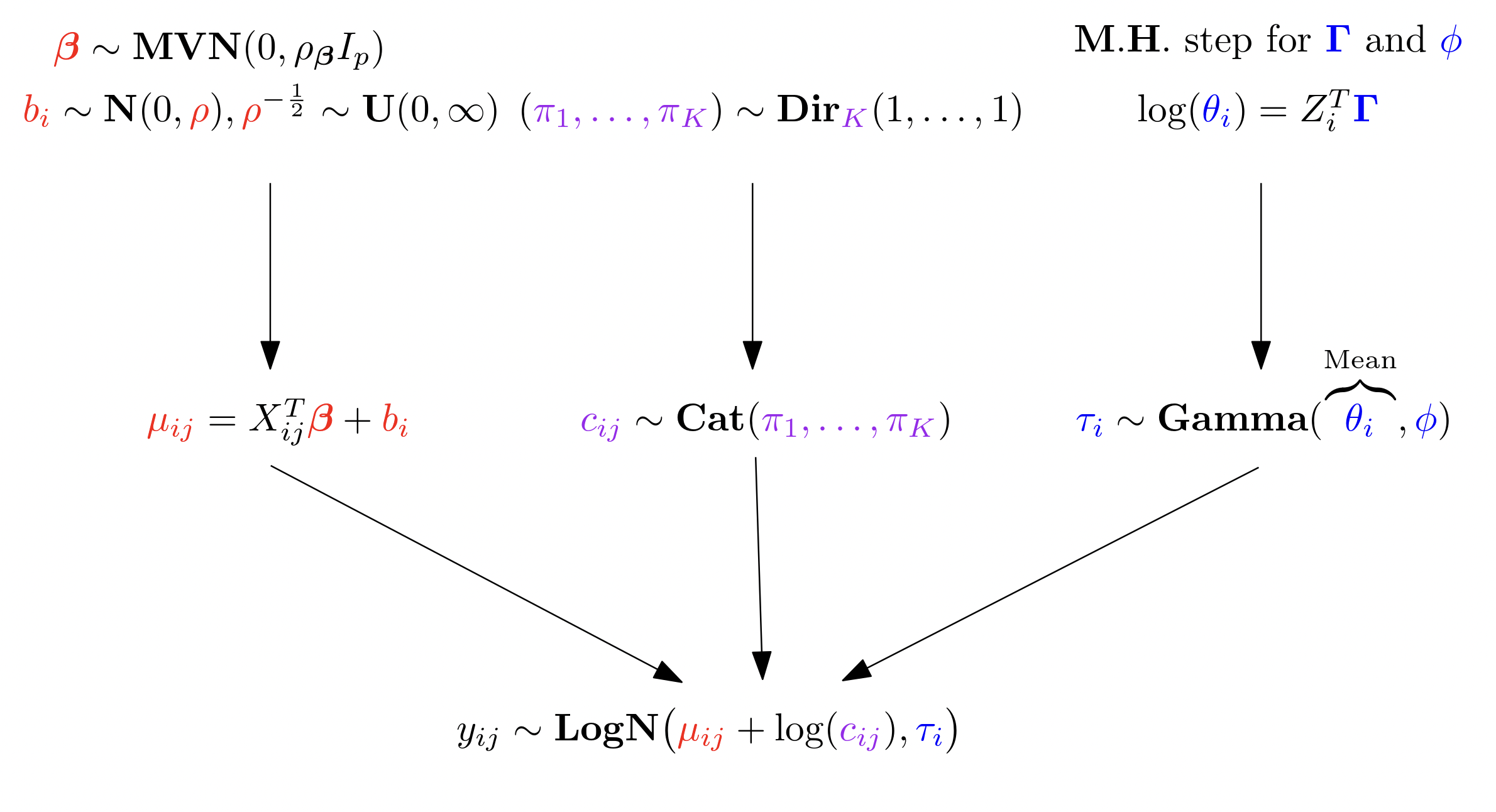}
    \caption{The SkipTrack Bayesian Hierarchical Model}
    \label{fig:skipTrackMod}
\end{figure}

In order to accurately model the right-skew of menstrual cycle length data (\cite{li2023menstrual}), we use the log-normal distribution for the observed lengths with the relation

\[
y_{ij} \sim \mathbf{LogNormal}\Big(\mu_{ij} + \log(c_{ij}), \tau_i\Big),
\]

\noindent where $\mu_{ij}$ is the logarithm of the expected median of the true cycle length when $c_{ij} = 1, \log(c_{ij}) = 0$, and $\tau_i$ is the precision. This parameterization allows for the separate investigation of an individual's cycle length, modeled with $\mu_{ij}$, and an individual's cycle `regularity,' modeled with $\tau_i$, while accounting for skips in self-tracking, modeled with $c_{ij}$. The three components of this model can be seen clearly in Figure \ref{fig:skipTrackMod}.

We include both baseline and longitudinal covariates for average cycle length by modeling a linear regression on $\mu_{ij}$  with an individual-specific random intercept component $b_i$ and a $p \times 1$ parameter vector $\boldsymbol\beta$. That is

\[
\mu_{ij} = X^T_{ij}\boldsymbol\beta + b_i,
\]

\noindent with priors with $b_i \sim \mathbf{Norm}(0, \rho)$, $\boldsymbol\beta \sim \mathbf{MVNorm}(0, \rho_{\boldsymbol\beta} \mathbf{I}_p)$ (both parameterized using precision), and $\rho^{-1/2} \sim \mathbf{Unif}(0, \infty)$.

Then we cap the maximum number of true cycles contained in an observed cycle considered by the model with $K$ and let $c_{ij} \sim \mathbf{Categorical}(\pi_1, \dots, \pi_K)$, where $\sum_{k}\pi_k = 1$ and $(\pi_1, \dots, \pi_K) \sim \mathbf{Dirichlet}_K(1, \dots, 1)$. The use of the categorical distribution here reduces restrictive assumptions on the probability of a skip in cycle length tracking allowing the data to better inform the results. While the model currently doesn't incorporate auxiliary information regarding the accuracy of an observed cycle, because the model is fit using MCMC sampling, $c_{ij}$ for a particular $i$ and $j$ can be fixed to a known value if desired.

Finally, we model the individual-level precision $\tau_i$ using baseline covariates as 

\[
\tau_i \sim \mathbf{Gamma}(\theta_i, \phi),
\]

\noindent where $\theta_i$ is the mean parameter and $\phi$ is the rate parameter. This allows for the inclusion of the covariates through the logarithm link-function with $\log(\theta_i) = Z_i^T\boldsymbol\Gamma$ where $\boldsymbol\Gamma$ is a $q \times 1$ parameter vector. 

We now discuss the algorithm for model estimation and inference.

\subsection{MCMC Sampling}

We use a multi-chain MCMC algorithm to sample from the SkipTrack model. Since most of the priors assumed in the model are conjugate or produce analytically solvable full conditional posteriors, most parameters in a single iteration of the MCMC algorithm are drawn with a Gibbs step (\cite{casella1992explaining}). The parameters drawn this way are $c_{ij}, \pi_k, \tau_i, b_i, \boldsymbol\beta, $ and $\rho$, for all $i,j,k$. The full conditional posteriors used in the Gibbs sampling steps on these parameters can be found in Appendix A.  In order to maintain an uninformative prior on $\boldsymbol\beta$ we set $\rho_{\boldsymbol\beta} = .01$.

The parameters $\boldsymbol\Gamma$ and $\phi$ cannot be drawn in a Gibbs step, and we instead use a Metropolis-Hastings step (\cite{chib1995understanding}). If $\boldsymbol\Gamma_t$ and $\phi_t$ are the current draws in the MCMC algorithm then we use proposal draws $\boldsymbol\Gamma^* \sim \mathbf{Normal}\big(\boldsymbol\Gamma_t, \rho_{\boldsymbol\Gamma}\big)$ and $\phi^* \sim \mathbf{Gamma}\big(\phi_t, \rho_\phi\big)$. This Normal distribution is parameterized using precision, and the Gamma is parameterized using mean and rate. Typically we set $\rho_{\boldsymbol\Gamma} = \rho_\phi = 1000$ to allow these parameters to change quickly over MCMC iterations.

\subsection{Computational Efficiency}\label{subsec: computationalEff}

The SkipTrack model has quite a few parameters and can be computationally burdensome when the number of individuals or cycles is large. In order to apply the SkipTrack model to large digitally-sourced datasets like the data from the AWHS, we apply the Wasserstein Posterior (WASP) method developed by \cite{srivastava2018scalable}.

First, the data is partitioned into $K$ subsets, ensuring that all cycles for a given individual are contained in exactly one of the subsets and the subsets all contain approximately the same number of individuals. Then, the SkipTrack model is run on each subset individually, resulting in $K$ posterior distributions for the cycle length and regularity parameters, denoted $\Pi_1(\boldsymbol\theta), \dots, \Pi_K(\boldsymbol\theta)$ and for $\boldsymbol{\theta} = (\boldsymbol\beta, \boldsymbol\Gamma)$. Finally, the WASP $\Pi_n(\boldsymbol\theta)$ is estimated using Algorithm 1 from \cite{srivastava2018scalable}.

This approach allows the SkipTrack model to be easily applied to massive digital datasets without requiring extensive computational times and while maintaining asymptotic validity as demonstrated in \cite{srivastava2018scalable}. We now explore the performance of the SkipTrack model in various situations through a simulation study.

\section{Simulations}\label{sec: Sims}

We use three simulation scenarios to evaluate the performance of the SkipTrack model. For each scenario we generate 800 datasets, 200 each for the number of individuals $n = 100, 500, 1000, 5000$, and present results for each sample size.

In Simulation 1, data is generated from the SkipTrack model with $n$ individuals. For each individual we generate three baseline covariates from a Gaussian distribution and use all three in $\mathbf{X}$ and $\mathbf{Z}$, as we wish to evaluate the performance of the model when covariates can effect both cycle length and regularity (see \cite{li2023menstrual}). Specific parameter settings may be found in the Supplementary Material (Duttweiler (2025)).

In Simulation 2, data is generated from the model found in \cite{li2022predictive} which is slightly modified to incorporate covariates affecting cycle length. For each individual, we generate three baseline covariates to effect the log-mean from a Gaussian distribution as in Simulation 1. Importantly, note that the use of the Poisson distribution in this model does not allow for the inclusion of precision parameters. The remainder of the specific parameter settings may be found in the Supplmentary Material (Duttweiler (2025)).

In Simulation 3, data is generated from a non-linear mixture model in order to examine model performance in the presence of a larger number of covariates and non-ideal circumstances. For each individual we generate 20 baseline covariates from a Gaussian distribution and set parameter values to effect cycle length mean and regularity. Specific details can be found in the Supplementary Material (Duttweiler (2025)).

\subsection{Skip-Tracking Accuracy}

We evaluate the performance of the SkipTrack model against the model presented in \cite{li2022predictive} in terms of accuracy in identifying user-skips in cycle tracking. Skips are estimated from the Li model by estimating the hyperparameters as discussed in the paper, and then using a Gibbs sampler based MCMC algorithm to sample from the posterior distribution of the $c_{ij}$ values. In each simulation, for both models, 5 chains were run for 10,000 iterations each with a burn-in of 750 draws (as convergence was reached quite quickly).

For both models we set $\hat{c}_{ij}$ to be the maximum a posteriori (MAP) estimator for each $i,j$ pair. In all simulations we observed an accuracy rate for $c_{ij} = \hat{c}_{ij}$ of around 96-98\% in both models, showing that both are highly capable of identifying skips. However, here we note an important difference that has implications for inference on the covariates. 

\begin{figure}[h]
    \centering
    \includegraphics[width=\linewidth]{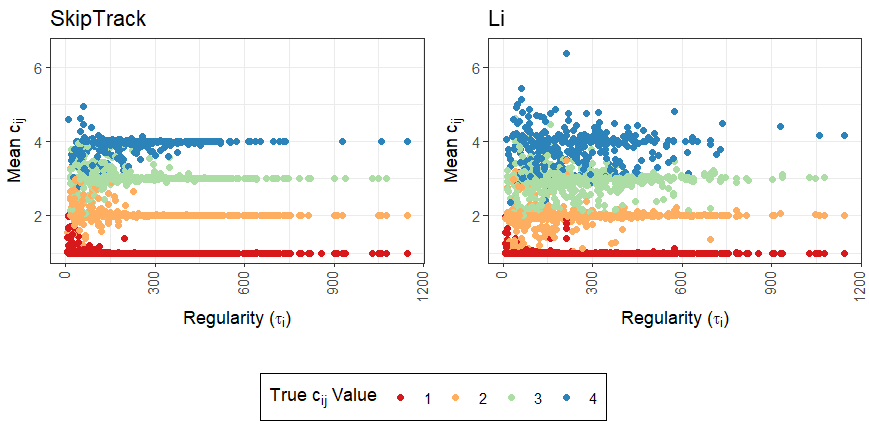}
    \caption{True regularity vs. posterior $c_{ij}$ means with the SkipTrack and Li models. The colors represent the true $c_{ij}$ values.}
    \label{fig: cijMeans}
\end{figure}

Figure \ref{fig: cijMeans} plots the underlying true $\tau_i$'s, representing individual $i$'s regularity, against each model's Monte Carlo averaged $c_{ij}$'s for one dataset from Simulation 1 with $n = 1000$. This demonstrates that, for the SkipTrack model, when $\tau_i$ is small (an individual with low regularity), the model is less certain when identifying skips. But, as $\tau_i$ grows the model quickly becomes confident in skip identification. This contrasts with the Li model, which shows uncertainty in skip identification while $\tau_i$ is small, and remains uncertain until $\tau_i$ is much larger. 

\begin{figure}[h]
    \centering
    \includegraphics[width=\linewidth]{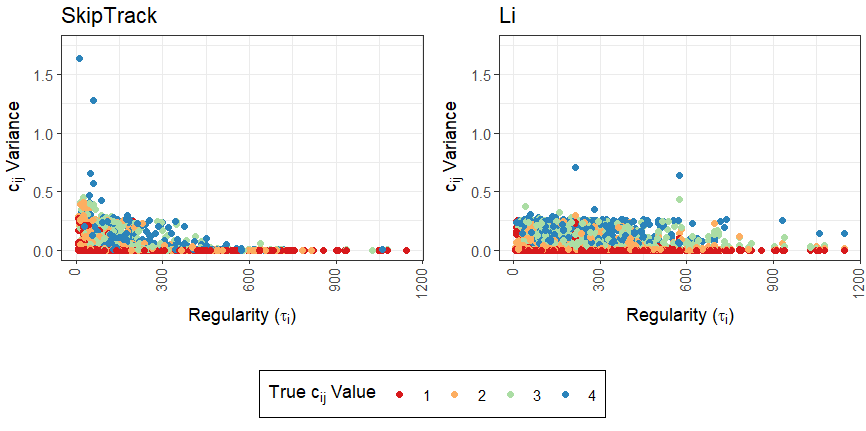}
    \caption{True regularity vs. posterior $c_{ij}$ variances with the SkipTrack and Li models. The colors represent the true $c_{ij}$ values.}
    \label{fig: cijVars}
\end{figure}

We see a similar effect in Figure \ref{fig: cijVars} which shows the underlying true $\tau_i$'s against each model's Monte Carlo variances for the $c_{ij}$'s in the same dataset. For the Li model, individual's underlying regularity appears (as expected) to make very little impact on the variance in the sampled $c_{ij}$ values. In contrast, for the SkipTrack model, $\tau_i$ and the variance in the sampled $c_{ij}$ values appear to have an inverse, possibly exponential, relationship.

Taken together, Figures \ref{fig: cijMeans} and \ref{fig: cijVars} demonstrate an important difference between the SkipTrack and the Li models. By accounting for individual regularity the SkipTrack model appropriately adapts its level of confidence in estimation of skips in tracking. In estimation and inference on the effects of covariates this will have the effect of allowing observations that are more likely to be accurate to be more influential than those that are less likely to be accurate. This is as opposed to the Li model which does not adjust for regularity, and so gives all observations the same weight.

This leads us to a conclusion that is strongly in favor of the SkipTrack model. Namely, utilizing the SkipTrack model limits the need to filter out individuals with irregular cycle lengths as the model will properly adjust for unstable observations on its own.

\subsection{Covariate Inference}

Now, we compare two models regarding inference for covariate effects on cycle length and regularity. First, the SkipTrack model is as presented. For the second, we simulate an approach, common in practice, in which skips in cycle tracking are estimated and then \textit{fixed} by estimating cycle skips using the MAP estimator from the model in \cite{li2022predictive} as described above. We then perform inference with the SkipTrack model while fixing the cycle skips to those estimated. 

\begin{table}[t]
    \centering
\begin{tabular}{ |c|c||c|c|c||c|c|c|}
 \hline
 \multicolumn{2}{|c||}{} & \multicolumn{3}{|c||}{SkipTrack} & \multicolumn{3}{|c|}{SkipTrack w/ Fixed Skips}\\
 \hline
  &$n$ & Bias & Width & Coverage & Bias & Width & Coverage \\
  \hline \hline
  \multirow{4}{5em}{$\beta_1 = -0.02$} 
  & 100 & 0.001 & 0.055 & 0.954 & 0.004 & 0.050 & 0.954 \\
  & 500 & 0.001 & 0.024 & 0.934 & 0.004 & 0.022 & 0.856 \\
  &1000 & 0.000 & 0.017 & 0.959 & 0.003 & 0.015 & 0.876 \\
  &5000 & 0.000 & 0.008 & 0.970 & 0.003 & 0.007 & 0.610 \\
  \hline
  \multirow{4}{5em}{$\beta_2 = 0.00$}
  & 100 & 0.001 & 0.056 & 0.919 & 0.001 & 0.050 & 0.944 \\
  & 500 & 0.000 & 0.024 & 0.949 & 0.000 & 0.022 & 0.964 \\
  &1000 & 0.000 & 0.017 & 0.954 & 0.000 & 0.015 & 0.953 \\
  &5000 & 0.000 & 0.008 & 0.960 & 0.000 & 0.007 & 0.940 \\
 \hline
 \multirow{4}{5em}{$\beta_3 = 0.06$}
 & 100 & 0.002 & 0.056 & 0.970 & -0.008 & 0.050 & 0.944 \\
 & 500 & 0.000 & 0.024 & 0.964 & -0.010 & 0.022 & 0.595 \\
 &1000 & 0.001 & 0.017 & 0.929 & -0.009 & 0.015 & 0.361 \\
 &5000 & 0.000 & 0.008 & 0.945 & -0.009 & 0.007 & 0.000 \\
 \hline\hline
  \multirow{4}{5em}{$\Gamma_1 = 0.0$} 
  & 100 &-0.002 & 0.316 & 0.924 & 0.002 & 0.316 & 0.919 \\
  & 500 & 0.001 & 0.138 & 0.939 & 0.003 & 0.138 & 0.954 \\
  &1000 & 0.000 & 0.098 & 0.939 & 0.001 & 0.097 & 0.947 \\
  &5000 &-0.001 & 0.043 & 0.955 &-0.001 & 0.043 & 0.955 \\ 
  \hline
  \multirow{4}{5em}{$\Gamma_2 = -0.1$}
  & 100 &-0.002 & 0.317 & 0.909 & 0.021 & 0.316 & 0.909 \\
  & 500 &-0.002 & 0.138 & 0.939 & 0.020 & 0.138 & 0.918 \\
  &1000 & 0.000 & 0.098 & 0.954 & 0.022 & 0.097 & 0.882 \\
  &5000 & 0.001 & 0.043 & 0.925 & 0.023 & 0.043 & 0.480 \\
 \hline
 \multirow{4}{5em}{$\Gamma_3 = 0.3$}
 & 100 & -0.001 & 0.324 & 0.954 & 0.017 & 0.321 & 0.934 \\
 & 500 & 0.001 & 0.141 & 0.919 & 0.015 & 0.140 & 0.903 \\
 &1000 & 0.001 & 0.099 & 0.964 & 0.011 & 0.099 & 0.893 \\
 &5000 &-0.001 & 0.044 & 0.935 & 0.009 & 0.044 & 0.790 \\
 \hline
\end{tabular}
\caption{Bias, 95\% credible interval width, and coverage for the log-median and precision parameters in Simulation 1.}
\label{tab: sim1Both}
\end{table}

Table \ref{tab: sim1Both} gives the results on the $\boldsymbol\beta$ and $\boldsymbol\Gamma$ parameters for the covariates respectively affecting the log-median and precision in Simulation 1. Width and coverage results are based on the 95\% credible intervals derived from the MCMC samples. Importantly, note that fixing the skips appears to bias the estimates for log-median ($\boldsymbol\beta$) parameters with a true effect, attenuating them towards zero and resulting in a loss of coverage as $n$ increases. The results for the precision parameters $\boldsymbol\Gamma$ demonstrate a similar attenuating effect for $\Gamma_2$, but a bias away from zero for $\Gamma_3$. As we discuss further with the results from Simulation 3, we believe this is related to the fact that $\beta_2 = 0$ but $\beta_3 \neq 0$. While we chose relatively modest sample sizes for computational feasibility of repeated simulations, this attenuating effect appears particularly important in our motivating application involving a large digital health dataset. 

The results from Simulation 2 follow the same pattern for the log-mean parameters $\boldsymbol\beta$, despite the fact that we use the Li model both to generate the data and to estimate the fixed skips. Table \ref{tab: sim2Betas} gives the results for the mean parameters in Simulation 2. The estimate attenuation when fixing skips is not as bad as in Simulation 1, but it still results in a loss of coverage as $n$ grows.

\begin{table}[h]
\centering
\begin{tabular}{|c|c||c|c|c||c|c|c|}
 \hline
 \multicolumn{2}{|c||}{} & \multicolumn{3}{|c||}{SkipTrack} & \multicolumn{3}{|c|}{SkipTrack w/ Fixed Skips}\\
 \hline
  &$n$ & Bias & Width & Coverage & Bias & Width & Coverage \\
  \hline \hline
  \multirow{4}{5em}{$\beta_1 = -0.02$} 
  & 100 & 0.001 & 0.045 & 0.965 & 0.002 & 0.042 & 0.970 \\
  & 500 & 0.000 & 0.020 & 0.960 & 0.001 & 0.018 & 0.960 \\
  &1000 & 0.000 & 0.014 & 0.935 & 0.001 & 0.013 & 0.940 \\
  &5000 & 0.000 & 0.006 & 0.975 & 0.001 & 0.006 & 0.915 \\
  \hline
  \multirow{4}{5em}{$\beta_2 = 0.00$}
  & 100 &-0.002 & 0.045 & 0.970 &-0.001 & 0.042 & 0.965 \\
  & 500 & 0.000 & 0.020 & 0.940 & 0.000 & 0.018 & 0.940 \\
  &1000 & 0.000 & 0.014 & 0.970 & 0.000 & 0.013 & 0.965 \\
  &5000 & 0.000 & 0.006 & 0.945 & 0.000 & 0.006 & 0.940 \\
 \hline
 \multirow{4}{5em}{$\beta_3 = 0.06$}
 & 100 & 0.000 & 0.046 & 0.930 &-0.003 & 0.042 & 0.920 \\
 & 500 & 0.001 & 0.020 & 0.955 &-0.003 & 0.018 & 0.925 \\
 &1000 & 0.001 & 0.014 & 0.950 &-0.002 & 0.013 & 0.879 \\
 &5000 & 0.000 & 0.006 & 0.935 &-0.003 & 0.006 & 0.575 \\
 \hline
\end{tabular}
\caption{Bias, 95\% credible interval width, and coverage for mean parameters in Simulation 2.}
\label{tab: sim2Betas}
\end{table}

When examining the results for Simulation 3, we use different criteria as the effects on cycle length mean are on a different scale than those expected by the SkipTrack model, and the effects on cycle regularity are non-linear. Instead of examining bias and coverage, we present the overall rates for both Type I (detecting an effect when none exists) and Type II (failing to detect an effect when one exists) errors. Table \ref{tab: sim3Errors} gives these error rates, separated by target and sample size.

First, note that the error rates for the mean parameters are essentially the same, whether fixing skips or not. This may appear to disagree with the results from Simulations 1 and 2, but in reality it does not, as the estimates on cycle length mean (or median) from those simulations were simply attenuated towards 0, not estimated to be 0. However, note the patterns of increased Type I error rate for the precision parameters. While the Type I error is inflated when both accounting for and fixing the skips, it appears that, as $n$ increases, the error when accounting for the skips decreases at a faster rate than when fixing skips. Although not shown here, this phenomenon is particularly pronounced for the covariates that have no effect on cycle regularity, but do have an effect on cycle length mean. It appears that not accounting for the variability in possible skipping may increase the rate of false discoveries for covariates affecting regularity, specifically when those covariates also effect the mean. 

\begin{table}[b]
    \centering
\begin{tabular}{|c|c||c|c||c|c|}
 \hline
 \multicolumn{2}{|c||}{} & \multicolumn{2}{|c||}{SkipTrack} & \multicolumn{2}{|c|}{SkipTrack w/ Fixed Skips}\\
 \hline
  &$n$ & Type I Error & Type II Error & Type I Error& Type II Error \\
  \hline \hline
  \multirow{4}{5em}{Mean Parameters} 
  & 100 & 0.050 & 0.639 & 0.049 & 0.637 \\
  & 500 & 0.040 & 0.274 & 0.044 & 0.276 \\
  &1000 & 0.043 & 0.187 & 0.050 & 0.186 \\
  &5000 & 0.048 & 0.066 & 0.056 & 0.066 \\
  \hline
  \multirow{4}{5em}{Precision Parameters}
  & 100 & 0.324 & 0.452 & 0.308 & 0.479 \\
  & 500 & 0.140 & 0.160 & 0.186 & 0.160 \\
  &1000 & 0.071 & 0.113 & 0.170 & 0.113 \\
  &5000 & 0.008 & 0.064 & 0.121 & 0.058 \\
 \hline
\end{tabular}
\caption{Type I and Type II error rates for mean and precision parameters in Simulation 3.}
\label{tab: sim3Errors}
\end{table}

\begin{figure}[t]
    \centering
    \includegraphics[width = .9\linewidth]{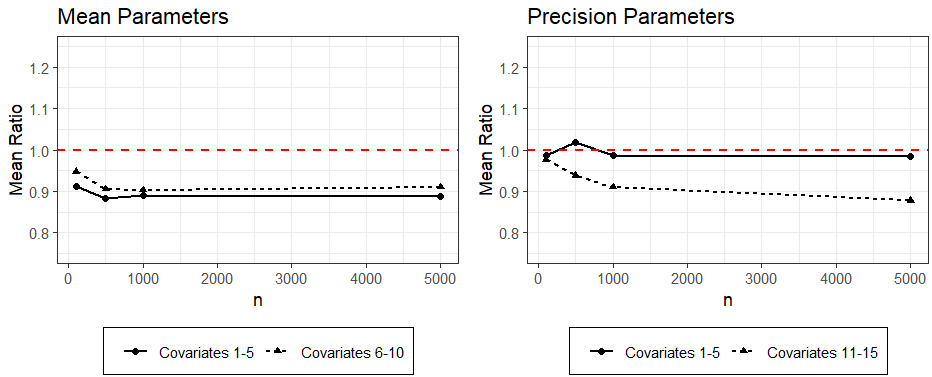}
    \caption{Ratio between estimates for model with fixed skips vs. model without fixed skips. }
    \label{fig: sim3Attn}
\end{figure}

In Figure \ref{fig: sim3Attn}, we further explore the estimate attenuation effect we had seen in Simulations 1 and 2. Let $\hat\beta_i$ be the estimate of $\beta_i$ calculated from the full SkipTrack model, and let $\hat\beta_i^F$ be the estimate of $\beta_i$ from the SkipTrack model after estimating and then fixing the skips. Figure \ref{fig: sim3Attn} presents $\hat\beta_i^F/\hat\beta_i$ (and $\hat\Gamma_i^F/\hat\Gamma_i$) averaged over simulation and covariate group, providing an estimate of effect attenuation caused by fixing the skip values instead of incorporating them in the model. We use Covariates 1-5 and 6-10 for the mean parameters, and Covariates 1-5 and 11-15 for the precision parameters as these are the non-zero coefficients for each set.

Notice that the mean parameter estimates for the fixed skips model are all about 90\% of the estimates obtained from the full model, supporting our results from Simulations 1 and 2. Interestingly, the precision parameter estimates are similarly attenuated for Covariates 11-15, when the true mean parameters are 0, but \textbf{not} attenuated for Covariates 1-5, when the true mean parameters are non-zero. This supports the results from Table \ref{tab: sim1Both}, which show a very slight bias away from zero for the estimate $\Gamma_3 = .3$, when $\beta_3 = .06$, but an attenuation towards zero for $\Gamma_2 = -.1$ when $\beta_2 = 0.$

These simulations demonstrate that the SkipTrack model is highly successful when estimating effects on both cycle length mean and regularity, whether dealing with possible skips in cycle tracking \textit{a priori}, or within the model. However, it is most accurate when allowing the model to account for skips as specifying skips \textit{a priori} can cause biased estimates or artificially inflate confidence by reducing the width of credible intervals.

We now apply the full SkipTrack model to the AWHS dataset described in Section \ref{sec: Data}.

\section{Application}\label{sec: Application}

As mentioned in Section \ref{sec: Data}, our primary goal in this analysis was to identify patterns of association between age, race/ethnicity and BMI, and cycle length and regularity. We utilized a dataset derived from the full AWHS data introduced in Section \ref{sec: Data}. Cycles were excluded from the analysis if during enrollment a participant reported being menopausal, using hormones, having a hysterectomy, having a PCOS, or having uterine fibroids, or the participant reported pregnancy or lactation during the cycle. Cycles were also excluded for missingness in any of these exclusion criteria. Additionally, participants with missing or biologically unreasonable values for age, race/ethnicity or BMI were excluded. All other covariates were treated with a missingness indicator. Exclusion details are visualized in Supplementary Figure 1. Finally, following the analyses in \cite{li2022predictive} and \cite{li2023menstrual}, we removed cycles outside of the 10-90 day range.

\begin{figure}[h]
    \centering
    \includegraphics[width=0.6\linewidth]{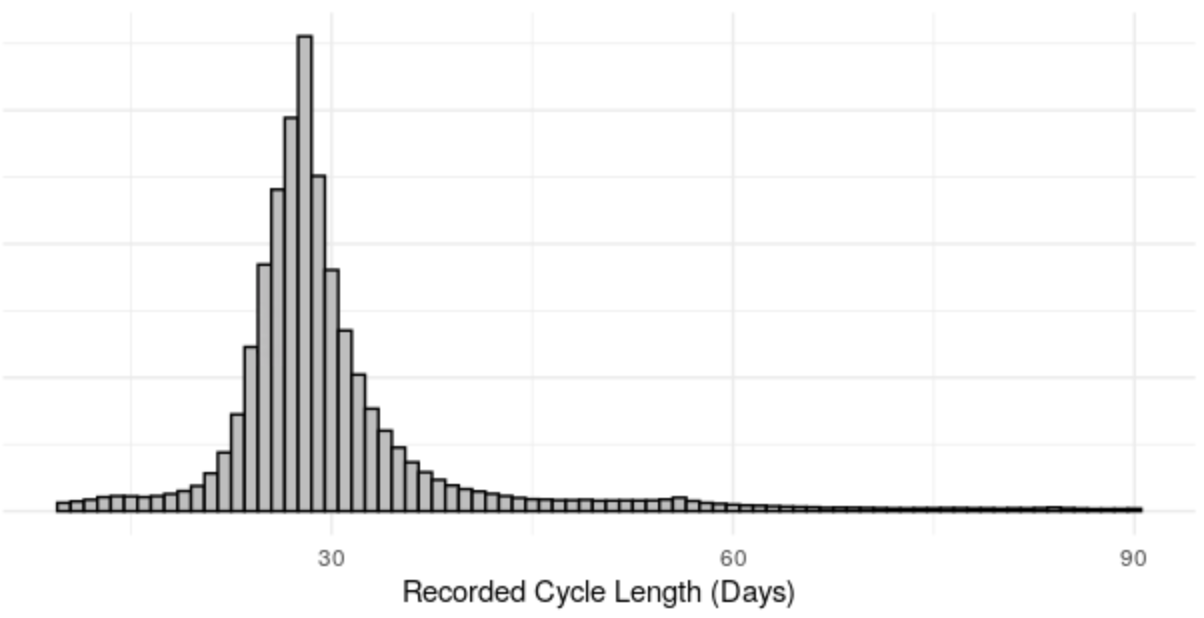}
    \caption{Histogram of recorded menstrual cycle lengths in the Apple Women's Health Study.}
    \label{fig: cycleHist}
\end{figure}

The resulting dataset for analysis contained 664,461 cycles contributed by 43,683 individuals, with individuals contributing a median of 13 cycles. The median contributed cycle length was 28 days, and the mean was 30 days. Figure \ref{fig: cycleHist} gives a histogram of the recorded cycles in this dataset. As can be seen, the majority of recorded cycles center around the overall median cycle length of 28 days. However, there are two small but noticeable peaks in the histogram at 56 and 84 days (two and three times 28, respectively) that indicate some skips in cycle tracking. 

Due to the large number of cycles recorded in the dataset we utilized the WASP method as outlined in Section \ref{subsec: computationalEff}, partitioning the data into $K = 20$ subsets. Each subset contained either 2,184 or 2,185 individuals with all of their recorded menstrual cycles. The run-time for the SkipTrack model on each subset was approximately 10 hours. 


We examined the $\hat{c}_{ij}$ estimates where $\hat{c}_{ij}$ is calculated as the \textit{maximum a posteriori} (MAP) of the sampled values. While the majority of the recorded cycles are estimated represent a single menstrual cycle, there are around 4\% of cycles that our model identifies as possibly resulting from skips in tracking. Interestingly, while the median cycle length for $\hat{c}_{ij} = 1,2,3$ was 28, 56, and 83 respectively, the maximum cycle length for each group was 90 days. This demonstrates that, not only are the estimates behaving as expected, but our model still allows some long cycles to be considered as biologically plausible and doesn't just bin all long cycles into a skipped category. 

\begin{figure}[b]
    \centering
    \includegraphics[width=0.8\linewidth]{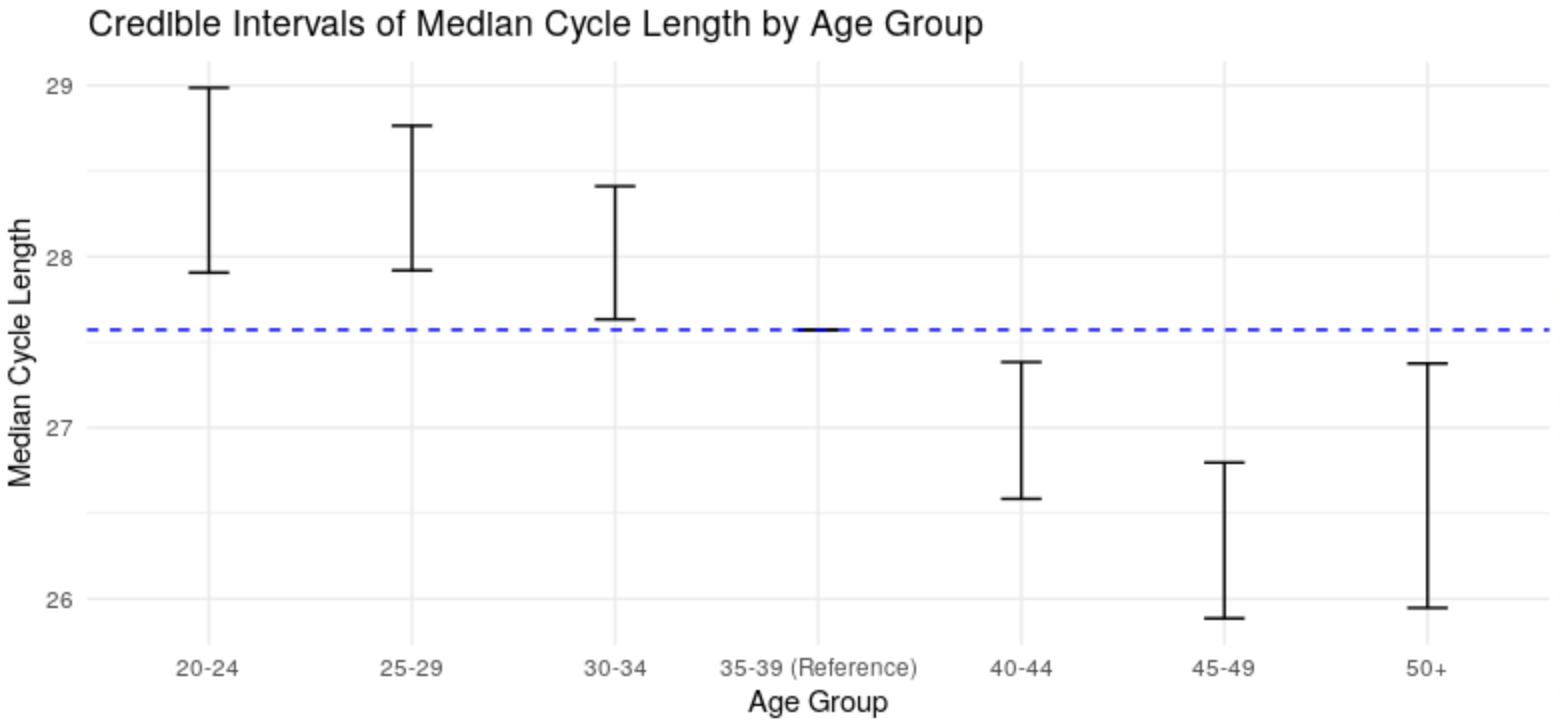}
    \caption{95\% credible intervals of median cycle length by age group, compared with reference group, age 35-39.}
    \label{fig: ageLength}
\end{figure}

Figure \ref{fig: ageLength} gives 95\% credible intervals for the estimated median cycle length by age group. There is a clear trend toward shorter cycles as age increases, with a movement back towards longer cycles for participants over 50 years old. This matches exposure-response patterns observed in other analyses and cohorts (see for example \cite{liu2004factors, li2023menstrual}).

\begin{figure}[h]
    \centering
    \includegraphics[width=0.8\linewidth]{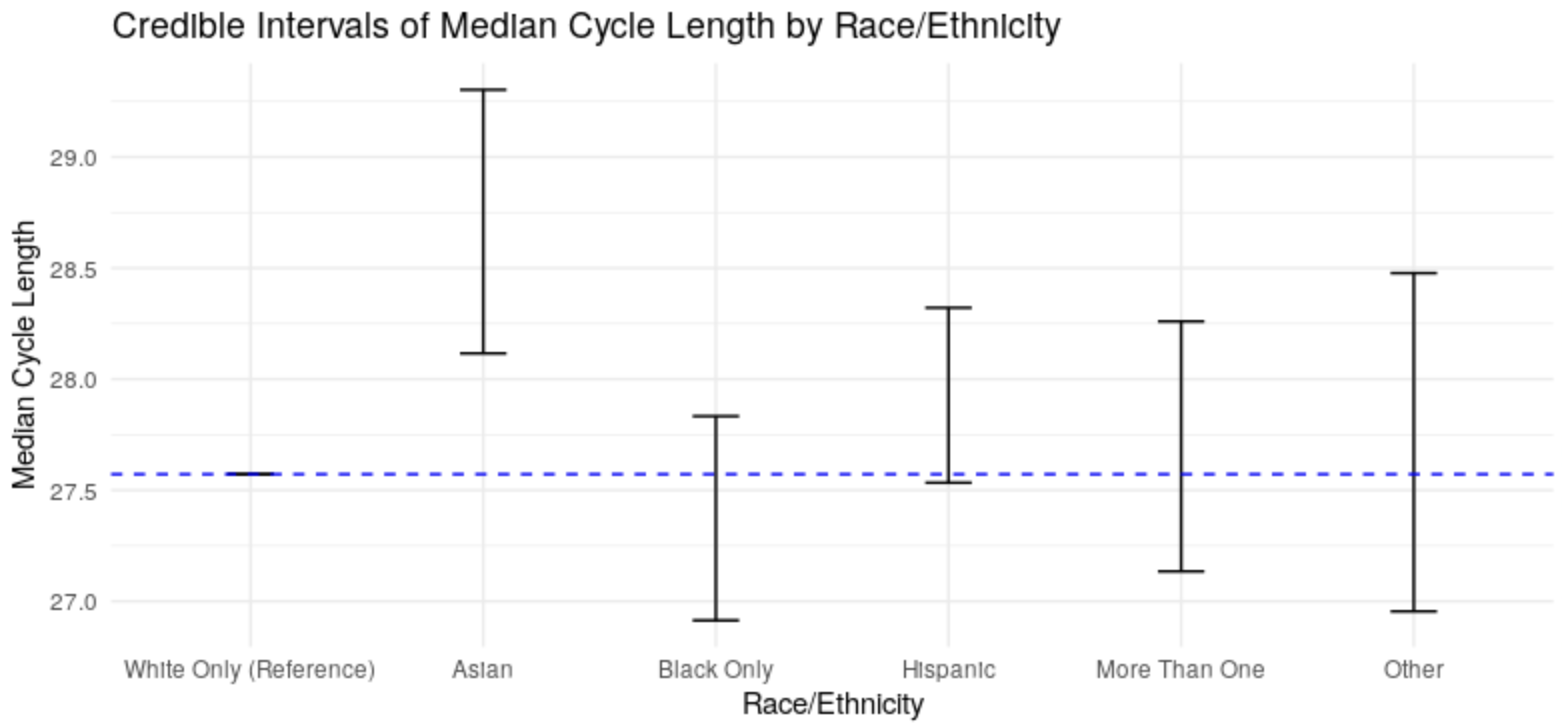}
    \caption{95\% credible intervals of median cycle length by race/ethnicity group, compared with reference group, White Only.}
    \label{fig: raceLength}
\end{figure}

Figure \ref{fig: raceLength} shows the 95\% credible intervals for the estimated median cycle length by race/ethnicity group. Similar to other studies, (see \cite{liu2004factors, paramsothy2015influence}) compared with White participants we observe longer cycle lengths for participants in the Asian category. Although the credible interval still contains the intercept, we also observe longer median cycle lengths for Hispanic participants, and shorter cycle lengths for Black participants. In our study, we observed no difference for individuals who recorded more than one race/ethnicity or who answered `Other'. 

\begin{figure}[h]
    \centering
    \includegraphics[width=0.8\linewidth]{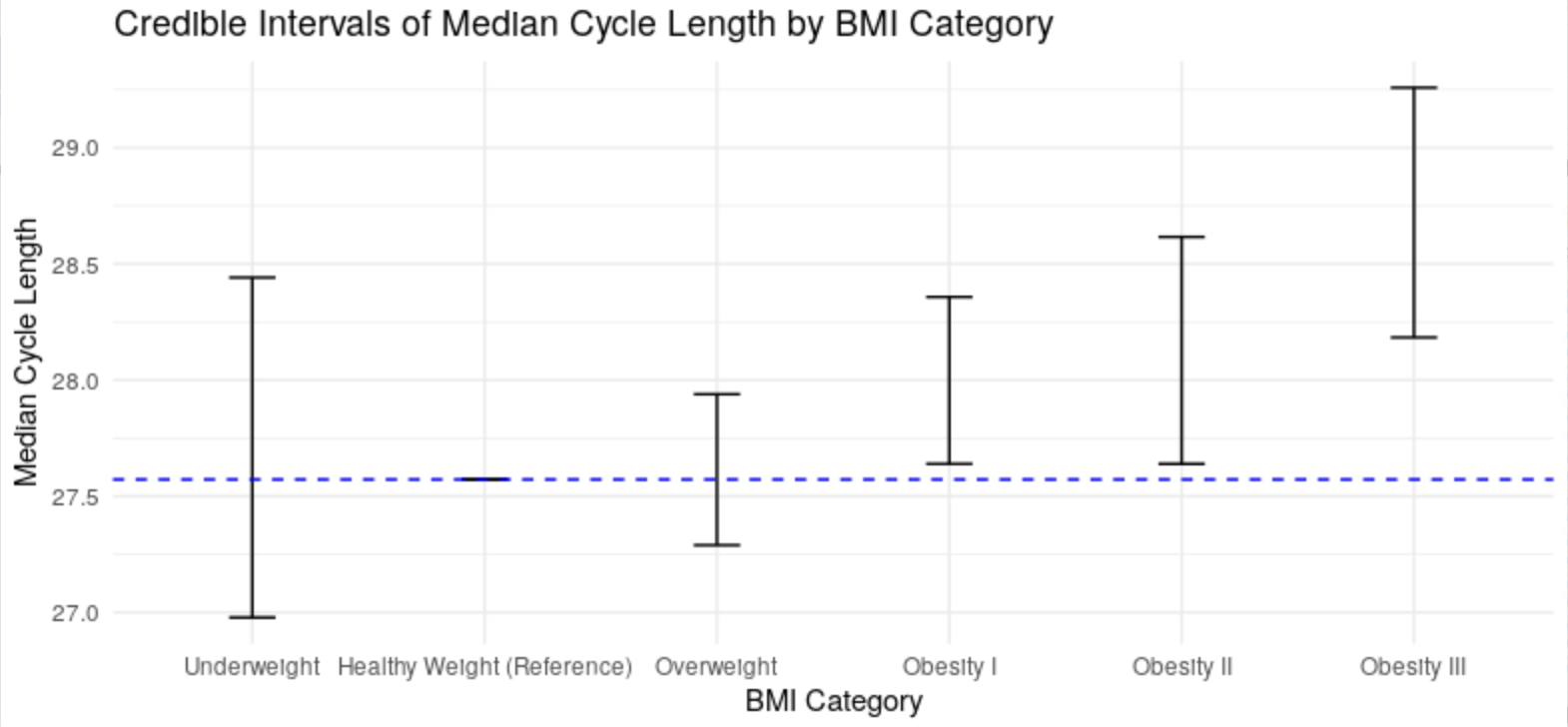}
    \caption{95\% credible intervals of median cycle length by BMI category, compared with reference group, Healthy Weight.}
    \label{fig: bmiLength}
\end{figure}

Figure \ref{fig: bmiLength} gives the 95\% credible intervals for the estimated median cycle length by BMI category. We observe longer median cycle lengths for individuals in the Obesity categories, again consistent with some established literature, although results on BMI have not been consistent (see for example \cite{liu2004factors, paramsothy2015influence, li2023menstrual}). 

While all but one of the age group credible intervals for regularity did contain the intercept, our estimates of effect of age on cycle regularity followed known patterns, with regularity increasing with age until 40-45 and then decreasing. In our analysis, no race/ethnicity categories showed a difference of cycle regularity compared against participants in the `White Only' category. Finally, all of the BMI categories for cycle regularity contained the intercept when compared against the `Healthy Weight' category. However, there is an observable decrease in the average regularity as BMI increases which may support the analysis done in \cite{li2023menstrual}. Figures showing the credible intervals for the effects of age, race/ethnicity, and BMI on regularity are contained in the Supplementary Material. 

\section{Discussion}\label{sec: Discussion}

In this paper we introduce the SkipTrack model for analyzing large digitally-based menstrual cycle cohorts. This approach improves on existing methods by placing the analysis of cycle length and regularity into a single model that additionally accounts for possible skips in cycle tracking, therefore reducing the estimation bias and overconfidence that result from \textit{a priori} removal of cycles with possible skips in tracking and exclusion of individuals with irregular cycle lengths. By incorporating the WASP divide-and-conquer approach, we are also able to scale the SkipTrack model to massive datasets, an important attribute for a analysis method for digital-health. The SkipTrack model is publicly available for R in the CRAN package \texttt{skipTrack} [\cite{duttweiler2024skipTrack, duttweiler2024skiptrackJOSS}]. 

An important limitation of this model is that skips in cycle tracking are essentially an `invisible' missing data problem. That is, recorded cycles that contain two or three true biological cycles are only recorded as such as we are missing data (user-recorded period bleeding days) that we cannot know are missing. This is a very difficult issue to work around, although some groups are approaching this problem creatively, by allowing users to designate cycles as `unusual' or by recording user-app interaction [\cite{li2020characterizing}].

Clearly established patterns in studies of menstrual cycle length and regularity show non-linear effects of various demographic and exposure variables. Although the SkipTrack model as stated above does not estimate non-linear effects in continuous variables, it could easily do so by incorporating penalized splines which have a natural Bayesian interpretation (see \cite{ruppert2003semiparametric}). In future work we seek to extend the SkipTrack model to account for non-linear effects, and to allow for the incorporation of information, such as sensor-based heart rate, respiratory rate or temperature measurements, that may improve our estimates for tracking skips. 

Our analysis of the Apple Women's Health Study data largely was in concordance with the wider literature on menstrual cycle length and regularity. Observed differences may be a result of our wider credible intervals that result from incorporating cycle length, regularity, and skips in tracking all into a single model, therefore fully propagating uncertainty.  

\nocite{duttweiler2025supplement}

\bibliographystyle{imsart-nameyear} 
\bibliography{bibliography}       


\end{document}